\documentclass[aps,prl,twocolumn,showpacs,superscriptaddress]{revtex4}
\usepackage{graphicx}
\usepackage{epstopdf}

\begin{document}

\title{Spatial Inhomogeneity of the Superconducting Gap and Order Parameter in FeSe$_{0.4}$Te$_{0.6}$}

\author{U. R. Singh}
\affiliation{Max-Planck-Institut f\"ur Festk\"orperforschung, Heisenbergstr. 1, D-70569 Stuttgart, Germany}
\author{S. C. White}
\affiliation{Max-Planck-Institut f\"ur Festk\"orperforschung, Heisenbergstr. 1, D-70569 Stuttgart, Germany}
\author{S. Schmaus}
\affiliation{Max-Planck-Institut f\"ur Festk\"orperforschung, Heisenbergstr. 1, D-70569 Stuttgart, Germany}
\author{V. Tsurkan}
\affiliation{Center for Electronic Correlations and Magnetism, Experimental Physics V, University of Augsburg, D-86159 Augsburg, Germany}
\affiliation{Institute of Applied Physics, Academy of Sciences of Moldova, MD 2028, Chisinau, R. Moldova}
\author{A. Loidl}
\affiliation{Center for Electronic Correlations and Magnetism, Experimental Physics V, University of Augsburg, D-86159 Augsburg, Germany}
\author{J. Deisenhofer}
\affiliation{Center for Electronic Correlations and Magnetism, Experimental Physics V, University of Augsburg, D-86159 Augsburg, Germany}
\author{P. Wahl}
\email{gpw2@st-andrews.ac.uk}
\affiliation{Max-Planck-Institut f\"ur Festk\"orperforschung, Heisenbergstr. 1, D-70569 Stuttgart, Germany}
\affiliation{SUPA, School of Physics and Astronomy, University of St. Andrews, North Haugh, St. Andrews, Fife, KY16 9SS, United Kingdom}

\date{\today}

\begin{abstract}
We have performed a low temperature scanning tunneling microscopy and spectroscopy study of the iron chalcogenide superconductor
$\mathrm{FeSe_{0.4}\mathrm{Te}_{0.6}}$ with $T_{\mathrm C}\approx 14~{\mathrm K}$. Spatially resolved measurements of the superconducting gap reveal substantial inhomogeneity on a nanometer length scale. Analysis of the structure of the gap seen in tunneling spectra by comparison with calculated spectra for different superconducting order parameters ($s$-wave, $d$-wave, and anisotropic $s$-wave) yields the best agreement for an order parameter with anisotropic $s$-wave symmetry with an anisotropy of $\sim 40\%$. The temperature dependence of the superconducting gap observed in places with large and small gap size indicates that it is indeed the superconducting transition temperature which is inhomogeneous. The temperature dependence of the gap size is substantially larger than would be expected from BCS theory. An analysis of the local gap size in relation with the local chemical composition shows almost no correlation with the local concentration of Se-/Te-atoms at the surface.
\end{abstract}


\maketitle
The recently discovered iron-based superconductors have sparked hope that a detailed understanding of superconductivity in these materials might
finally help to establish an understanding of the pairing mechanism in high temperature superconductors \cite{Stewart, Hirschfeld, Chubukov}.
The observation of magnetic resonance modes at the nesting vector of different Fermi surface sheets indicates that spin fluctuations play an
important role for superconductivity in these materials \cite{Qiu, Inosov, Mazin}. However despite these successes, there is still a number of open
questions to be resolved. The symmetry of the superconducting order parameter has not been unambiguously determined so far, also a predictive theory of superconductivity in iron-based superconductors is still missing. Matters are complicated by a complex band structure with up to five bands derived from the Fe-3d orbitals crossing the Fermi level \cite{Hirschfeld}. In the iron chalcogenide superconductor $\mathrm{Fe}_{1+\delta}\mathrm{Se}_{1-x}\mathrm{Te}_x$ it appears that the superconducting gap observed in tunneling spectra near optimal doping ($x\approx 0.6$) is nodeless \cite{Hanaguri} - while in MBE-grown FeSe films, it appears to have nodes \cite{Song}. An anisotropy of the superconducting gap has been observed in studies of $\mathrm{LiFeAs}$ by scanning tunneling microscopy (STM) \cite{Allan} and angle resolved photoemission (ARPES) \cite{Umezawa}. In the case of $\mathrm{Fe}_{1+\delta}\mathrm{Se}_{1-x}\mathrm{Te}_x$, results from ARPES experiments have been inconclusive: both, isotropic gaps on the hole-like and electron-like sheets of the Fermi surface - though of different magnitude \cite{Miao}, as well as anisotropic gaps \cite{Okazaki} have been reported. The latter is consistent with angle-resolved specific heat measurements which show evidence for an anisotropic gap in this sample \cite{Zeng}. A quasiparticle interference study by STM indicates that the order parameter reverses sign between different sheets of the Fermi surface, supporting an interpretation in terms of an $s_\pm$ order parameter \cite{Hanaguri}. The superconducting gap has been found to be inhomogeneous in iron pnictide superconductors of the 122-family \cite{Yin, Shan-natphys}. However, in the 122 materials cleaving usually creates a disordered surface, so this inhomogeneity is likely not representative of the bulk.

In this letter, we report a study of the spatial inhomogeneity and structure of the superconducting gap in $\mathrm{FeSe}_{0.4}\mathrm{Te}_{0.6}$ by STM. The temperature dependence of the gap shows that the inhomogeneity and spatial variations of the transition temperature are closely related to each other. A comparison of the local variation of the superconducting gap size with the anion height reveals almost no correlation, indicating that interlayer coupling is not negligible.

The 11 iron-chalcogenide superconductors have the simplest crystal structure of the iron-based superconductors, consisting of planar iron
layers with chalcogenide ($\mathrm{Se}$, $\mathrm{Te}$) anions above and below. The crystal structure provides a well-defined and non-polar cleavage plane between the chalcogenide layers. LEED and STM studies show no indication for a surface reconstruction \cite{Tamai, Massee}. We have carried out STM measurements on a single crystal of $\mathrm{FeSe}_{1-x}\mathrm{Te}_{x}$ with $x=0.61$ (determined by EDX measurements) and a superconducting transition temperature $T_{\mathrm C}\approx 14~\mathrm{K}$ \cite{Tsurkan}. We have used a home-built low temperature STM which allows for in-situ sample transfer and cleavage \cite{White2011}. Differential tunneling conductance $\mathrm dI/\mathrm dV$ is measured through a lock-in amplifier with a modulation of $600~{\mu\mathrm V}_\mathrm{RMS}$. Bias voltages are applied to the sample, with the tip at virtual ground. Tunneling spectra are acquired with open feedback loop. Sample cleaving was performed at temperatures around $20~\mathrm{K}$.

\begin{figure}[t]
\includegraphics [width=8.5cm]{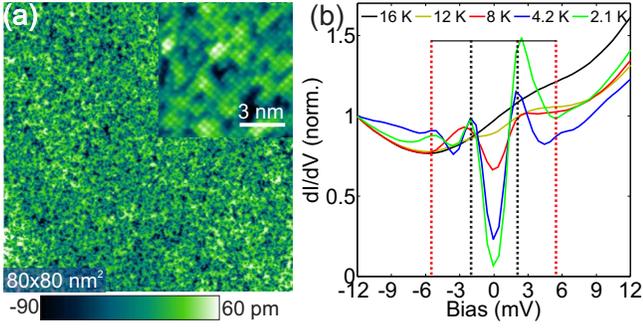}
\caption {(a) Atomically resolved topography of $\mathrm{FeSe}_{0.4}\mathrm{Te}_{0.6}$ ($80~\mathrm{nm}\times 80~\mathrm{nm}$), taken at $2.1~\mathrm{K}$ ($V= 90~\mathrm{mV}$, $I= 0.2~\mathrm{nA}$). Inset in (a) shows a magnified image of the atomic lattice, which contains $\mathrm{Se}$ and $\mathrm{Te}$ atoms. (b) Temperature dependent spectra ($4096$ single spectra) averaged over an area of size $20\times 20 ~\mathrm{nm}^2$ (stabilization condition: $V= 40~\mathrm{mV}, I= 0.5~\mathrm{nA}$).}
\label{fig:1}
\end{figure}
Figure \ref{fig:1}(a) shows a topographic image, the apparent inhomogeneity is dominated by the distribution of $\mathrm{Se}$ and $\mathrm{Te}$ atoms, yielding a larger height for $\mathrm{Te}$ atoms and a smaller height for $\mathrm{Se}$ atoms\cite{He2011}. A composition analysis based on the apparent height results in a tellurium concentration of $x=0.63\pm0.04$, consistent with EDX measurements. We detect almost no excess iron impurities. Temperature dependent spatially averaged $\mathrm dI/\mathrm dV$ spectra from spectroscopic maps are depicted in Fig. \ref{fig:1}(b), showing a superconducting gap similar to the one observed previously by Hanaguri {\it et al.}\cite{Hanaguri}. The superconducting gap is found to disappear roughly at $T_\mathrm C$. We note a pronounced asymmetry in the spectra, with different amplitude of the coherence peaks at positive and negative bias voltages. The asymmetry is position dependent. In addition to the dominant coherence peaks at $\pm 2.1~\mathrm{mV}$ (marked by dashed black lines), additional features can been seen outside the gap around $\pm 5.5~\mathrm{mV}$ (dashed red lines). These outer features are less reproducible than the inner coherence peaks, possibly because the orbital character of the associated bands couples only weakly to the tip of the STM. We can only speculate that they are likely due to a second, larger superconducting gap - however they are not exactly symmetric with respect to the Fermi energy. Therefore, we concentrate on the lower energy peaks in the following analysis.

\begin{figure}[!h]
\includegraphics [width=8.5cm]{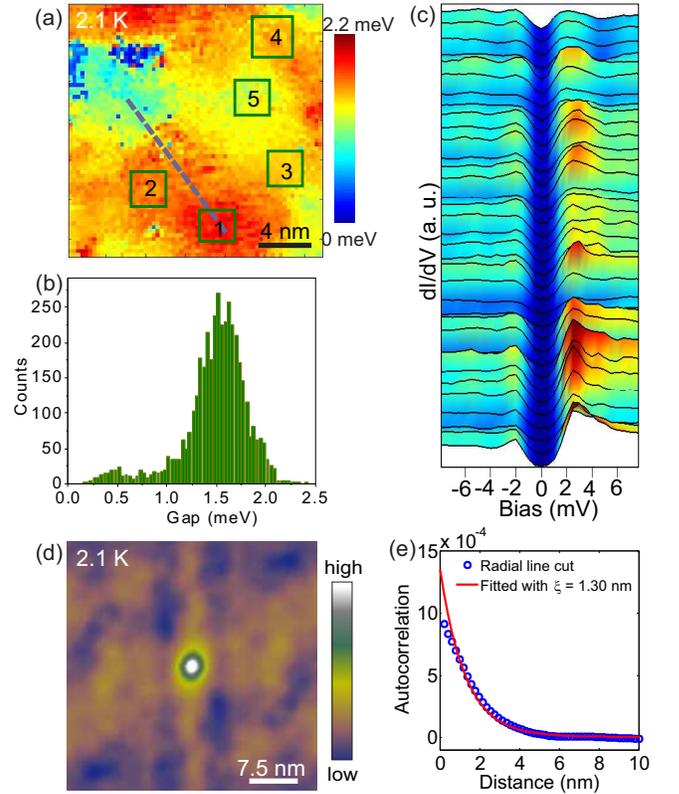}
\caption {(a) Gap map of $\mathrm{FeSe}_{0.4}\mathrm{Te}_{0.6}$ taken in an area of $20\times 20 ~\mathrm{nm}^2$ obtained from a spectroscopic map taken with a tunneling setpoint of $V= 40~\mathrm{mV}$ and $I= 0.5~\mathrm{nA}$ before switching off the feed-back loop. Over this area we have acquired temperature dependent spectroscopic maps with $64\times64$ points. (b) Histogram of the gap distribution. (c) Raw spectra along the line cut shown in (a). (d) Autocorrelation of the gap map taken over a large field of view ($38\times 38 ~\mathrm{nm}^2$). (e) radial line cut around the center of the autocorrelation shown in (d) obtained by averaging azimuthally (blue), the solid red line represents a fit of an exponential decay to the radial line cut.}
\label{fig:2}
\end{figure}
For the investigation of the spatial variation of the size of the superconducting gap, spectra in the map are fitted for one polarity with a
Dynes' gap function,
\begin{equation}
\frac{dI}{dV}(V)= \left|\mathrm{Re} \left[\frac{eV-i\Gamma}{\sqrt{(eV-i\Gamma)^2-\Delta^2}}\right]\right|,
\label{dynes-eq}
\end{equation}
where $\Delta$ and $\Gamma$ represent the size of the superconducting gap and quasiparticle-lifetime broadening, respectively. A spatial map of the local gap size obtained at positive polarity is presented in Fig. \ref{fig:2}(a) (the one obtained for negative polarity shows the same qualitative behaviour). It can clearly be seen that the size of the superconducting gap is spatially inhomogeneous with a characteristic length scale on the order of one nanometer and varies between $0.25~\mathrm{meV}$ and $2.2~\mathrm{meV}$ (see Fig.~\ref{fig:2}(b)). In Fig.~\ref{fig:2}(c), a series of spectra obtained along the line shown in Fig.~\ref{fig:2}(a) shows the coherence peaks at $\pm 2.1~\mathrm{meV}$ evolving from large gap (bottom) to small gap (top), where the coherence peaks almost disappear and rather only a depletion in the differential conductance is visible. A similar type of gap inhomogeneity has been found in the iron-arsenide compound $\mathrm{BaFe}_{1.8}\mathrm{Co}_{0.2}\mathrm{As}_{2}$ by STS and has been explained by impurity scattering \cite{Yin}. We detect very few impurities, therefore it is unlikely that this is the main cause of inhomogeneity in our measurements. Thus, the main source of inhomogeneity in our sample has to be the disorder of $\mathrm{Se}$/$\mathrm{Te}$ ions. However, in contrast to high temperature cuprate superconductors where the inhomogeneity arises due to disorder of the dopant atoms \cite{Pan-Lang}, here the substitution of $\mathrm{Se}$ by $\mathrm{Te}$ atoms is isoelectronic, so the mechanism linking the inhomogeneity to the local variation in the size of the superconducting gap has to be different.

To quantify the characteristic length scale of the inhomogeneity, which is also a measure for the coherence length, we have calculated the autocorrelation of the gap map (see Fig.~\ref{fig:2}(d)). It shows a slight anisotropy between the two nominally equivalent Fe-Fe bond direction, which can be rationalized by nematic excitations detected in the same crystal \cite{Singh2013}. By fitting an exponential decay function to the radially averaged line profile (see fig.~\ref{fig:2}(e)) we find a decay length of $\xi= 1.30~\mathrm{nm}$, in good agreement with the coherence length obtained from $H_{c2}$ of $1.5~\mathrm{nm}$ \cite{Klein}.

For a detailed analysis of the gap size as a function of temperature, we have compared our spectra with fits for different order
parameters in order to determine which yields the best description. To this end, we have introduced different angular dependent order parameters
$\Delta(\theta)$ into Eq.~\ref{dynes-eq}. We have considered the cases of pure s-wave ($\Delta(\theta) =\Delta_{0}$), anisotropic s-wave
($\Delta(\theta) =\Delta_{0}+\Delta_{1}\cos4\theta$), and d-wave ($\Delta(\theta) =\Delta_{0}\cos4\theta$) order parameters. Furthermore,
to describe the temperature dependence of the spectra, we have accounted for the thermal broadening of the Fermi function.

The differential conductance $\mathrm dI/\mathrm dV$ measured by STM can be considered proportional to the density of states of the sample, where the proportionality constant depends on the tip height and details of the tip apex \cite{Bardeen-Chen}. To eliminate these effects and contributions from the normal state density of states in our spectra taken below $T_{\mathrm C}$, we have divided these by spectra acquired with the same tunneling parameters in the normal state at $T>T_{\mathrm C}$.

\begin{figure}[t]
\includegraphics [width=5cm]{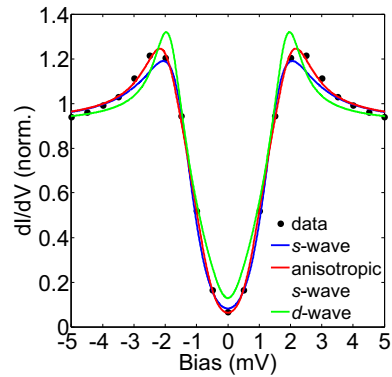}
\caption {The black symbols represent a spectrum extracted from a spatial average of the map shown in Fig. \ref{fig:2}(a) at $2.1~\mathrm{K}$.
The spectrum has been symmetrized and normalized as discussed in the text. Blue, red, and green solid lines are fitted curves using Eq.~\ref{dynes-eq} with different gap functions: $s$-wave ($\Delta(\theta)=\Delta_{0}$), anisotropic $s$-wave ($\Delta_{0}+\Delta_{1}\cos4\theta$), and $d$-wave ($\Delta_{0}\cos2\theta$), respectively.}
\label{fig:3}
\end{figure}
Figure~\ref{fig:3} shows a spatially averaged spectrum taken at $2.1~\mathrm{K}$, normalized and symmetrized around zero bias (symbols). The solid lines show Dynes equation (Eq.~\ref{dynes-eq}) fits with different order parameters $\Delta(\theta)$. The best fit is obtained for the anisotropic $s$-wave scenario - pure $s$-wave and $d$-wave do not give the same level of agreement. The extracted gap size is $\Delta_{0}= 1.42~\mathrm{meV}$ -- close to the results obtained from previous STM measurement showing coherence peaks in the tunneling spectra at $\pm 1.7~\mathrm{meV}$ \cite{Hanaguri} and ARPES \cite{Miao}, and somewhat smaller than what has been seen by optical spectroscopy ($2.5~\mathrm{meV}$) \cite{Homes}. For the anisotropy we obtain $\Delta_{1}= 0.60~\mathrm{meV}$. While we do fit the anisotropy by considering a $\cos 4\theta$ term, we cannot exclude that the anisotropy is governed, e.g., by a $\cos 2\theta$ term, because the resulting spectrum remains the same for any integer multiple of $\theta$.

The anisotropy which we obtain is about $\sim40\%$ of the gap magnitude, consistent with angle-resolved specific heat experiments for
samples of slightly different composition ($\mathrm{FeSe}_{0.45}\mathrm{Te}_{0.55}$) \cite{Zeng}, for which an anisotropy of $\approx 50\%$ was found. Recently, from measurements by Laser ARPES an anisotropy of $25\%$ has been reported with a superconducting gap of $\Delta_{0} = 1.63 ~\mathrm{meV}$ at $2.5~\mathrm{K}$ \cite{Okazaki}, both are quite close to the values extracted from our fits.

\begin{figure}[t]
\includegraphics [width=8.2cm]{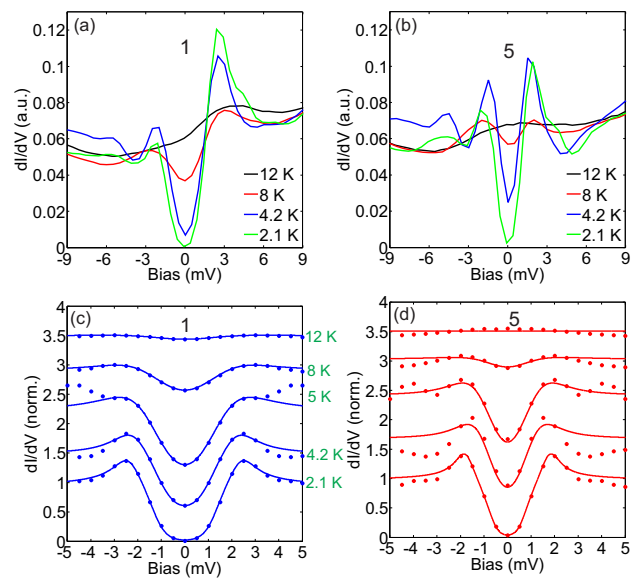}
\caption {(a) and (b) show averaged spectra acquired at different temperatures over two different areas of size
$\sim 3\times3~\mathrm{nm}^2$ each in regions with large (position 1 in Fig. \ref{fig:2}(a)) and small (position 5 in Fig. \ref{fig:2}(a)) local gap size. (c), (d) Symmetrized and normalized spectra taken at different temperatures in regions of large (a) and small (b)
gap size with fits (solid lines) according to Eq.\ref{dynes-eq} for an anisotropic s-wave order parameter and accounting for thermal broadening.
Spectra are normalized by spectra taken at $16~\mathrm{K}$. }
\label{fig:4}
\end{figure}
In order to explore the evolution of the superconducting gap with temperature we have acquired spectroscopic maps at temperatures between $2.1~\mathrm{K}$ and $12~\mathrm{K}$ in the same area of the sample as shown in Fig.~\ref{fig:2}(a). The temperature dependent spectra taken in regions with large and small local gap size (see Fig.~\ref{fig:4}(a) and (b)) reveal that in different regions the superconducting gap disappears at different temperatures. For example, spectra taken at $12~\mathrm{K}$ in a region with a large local gap size (Fig.~\ref{fig:4}(a)) still show a slight dip near zero-bias voltage which is absent in spectra taken in a region with a small gap size (Fig.~\ref{fig:4}(b)) at the same temperature.\\

\begin{figure}
\includegraphics [width= 8.7cm]{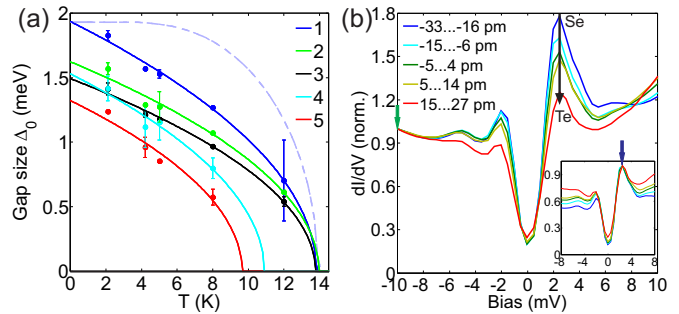}
\caption {(a) Gap size $\Delta_0$ as a function of $T$ for different gap regions (marked in Fig. \ref{fig:2} (a)), showing local differences in the temperature at which the gap is completely closed. Solid lines are fits of a $\sqrt{1-T/T_\mathrm C}$ behaviour for the temperature dependence of the gap size. The dashed light blue line is the temperature dependence of the superconducting gap as it would be expected from BCS theory for an $s$-wave gap (assuming the same $T_\mathrm C$ and $\Delta_0(0)$ as obtained from the fit of the dark blue line for region 1). (b) Tunneling spectra from a spectroscopic map averaged for different ranges of relative topographic heights and normalized at $V=-10~\mathrm{mV}$ (marked by green arrow). The same spectra normalized at the energy of the coherence peak (marked by blue arrow).}
\label{fig:5}
\end{figure}

We can determine local transition temperatures $T_\mathrm{C}$ by fitting Eq.~\ref{dynes-eq}, with the anisotropic s-wave gap function to spectra averaged over regions with similar local gap size. In Fig.~\ref{fig:4}(c) and \ref{fig:4}(d) we show symmetrized and normalized spectra from regions with large and small local gap size taken at different temperatures (note that some spectra show spurious features which are tip-related, see e.g. spectra taken at $5\mathrm K$ in Fig.~\ref{fig:4}(c, d) near $\pm5\mathrm{mV}$). The values of the gap size $\Delta_0$ obtained from the fits show a monotonic decrease with increasing temperature as shown in Fig.~\ref{fig:5}(a), they follow a $\sqrt{1-T/T_{\mathrm C}}$ behaviour. The temperature dependence differs from that expected from BCS theory for a weak coupling superconductor, especially at low temperatures $T<T_{\mathrm C}$, where the gap size becomes almost independent of temperature according to BCS theory \cite{Tinkham}. The critical temperatures which we obtain from regions of different gap size range from $10$ to $14~\mathrm K$. Thus, it is really the superconducting transition temperature which is spatially inhomogeneous in $\mathrm{FeSe}_{0.4}\mathrm{Te}_{0.6}$. From the fits, the gap size $\Delta_0(0)$ in the limit of low temperature can be extracted, yielding a ratio $2\Delta_0(0)/k_\mathrm BT_{\mathrm C}$ in a range from $2.5$ to $3.2$. This value is somewhat smaller than what would be expected from weak coupling BCS theory which gives $2\Delta_0(0)=3.52k_\mathrm BT_\mathrm C$.\\

Having established the relation between local gap size and transition temperature, we can compare the local superconducting gap size with the local chemical composition. As confirmed by X-ray diffraction, the chalcogen height of Se and Te atoms above the iron layer differs substantially \cite{Tegel}. We can extract a measure of the local chalcogen height from the apparent height in the topographic image acquired simultaneously with the map. In Fig.~\ref{fig:5}(b), tunneling spectra extracted from a map are averaged for different relative topographic heights, revealing that the height of the coherence peak decreases with the increase in the chalcogen height but the change in the superconducting gap size is very small. From spectra normalized at the coherence peak energy (see inset of Fig.~\ref{fig:5}(b)) it can be seen that the gap size is almost independent of anion height. The influence of the anion height on electronic, superconducting and magnetic properties of iron-based superconductors and their parent compounds has been widely discussed in literature \cite{Singh, Kuroki, Moon}. Details of the Fermi surface depend on the pnictogen/chalcogen height, for iron calcogenides, it has been shown that different magnetic orders are stabilized depending on the chalcogen height \cite{Moon}. Also its influence on superconductivity has been investigated \cite{Kuroki}, showing that even a change in the symmetry of the superconducting order parameter can occur as a function of anion height. This sensitivity stems from the chemical bond between iron $d$-orbitals and the chalcogen or pnictogen $p$-orbitals, whose strength depends strongly on the bond angle. The lack of a clear correlation between the local anion height and the size of the superconducting gap in our measurements indicates that there is substantial interlayer coupling and the superconducting properties are not only governed by the chemistry within one iron chalcogenide layer. DFT calculations comparing the band structures between FeSe and FeTe show that indeed in FeTe there are bands at the Fermi energy with a strong dispersion in the direction perpendicular to the iron chalcogenide planes \cite{Subedi}, suggesting substantial interlayer coupling, which plays a smaller role in FeSe. Our data suggest that at the doping of our sample the superconducting properties are already substantially influenced by coupling between the layers.\\

In conclusion, we have studied the spatial inhomogeneity and temperature dependence of the superconducting gap of $\mathrm{FeSe}_{1-x}\mathrm{Te}_{x}$ with $x=0.6$. We find that the spectra are best described by an anisotropic $s$-wave gap function with an anisotropy of $\sim 40\%$. Temperature dependent spectra acquired with atomic registry show that the local variation in gap size is directly linked to a local variation in the superconducting transition temperature. The correlation with the local concentration of selenium and tellurium atoms shows no clear trend. Our data indicate, that the local superconducting gap size is not only determined by the chemical composition within the top-most iron chalcogenide layer, but it is also influenced by deeper layers.

URS acknowledges support by the Alexander-von-Humboldt foundation and SCW by SPP1458 of the DFG. Support is also acknowledged via SPP 1458 through
project DE1762/1-1.\\

\end{document}